\documentstyle[epsfig]{qcdparis}
\def\stackunder#1#2{\mathrel{\mathop{#2}\limits_{#1}}}%
\begin{document}
\pagestyle{plain}
\title{
Unified QCD evolution equations for quarks and gluons
}

\author{
 S. Wallon}

\affil{ Service de Physique Th\'eorique, CE-Saclay\\

F-91191 Gif-sur-Yvette Cedex, FRANCE}

\abstract{Considering the BFKL and DGLAP QCD evolution equations for structure
functions,
we discuss the possibility of unifying them in the whole $x$ and
$Q^2$ range. We emphasize that the main problem is related to the constraint
of angular ordering of the radiation, and the cancellation of the related
collinear singularities for inclusive processes. At the leading log\ $1/x$ and
log\ $Q^2$ level, we write down a unified system of equations satisfying this
cancellation constraint. At low $x$, it leads to a less singular behaviour of
the structure functions than the BFKL prediction.}

\resume{Partant des \'equations dites BFKL et DGLAP de CDQ pour les fonctions
de
structure, nous discutons la possibilit\'e d'unifier celles-ci dans tout le
domaine en $x$ et $Q^2$. Nous montrons que la principale difficult\'e
vient de l'ordonnancement angulaire des radiations, et de l'annulation
des singularit\'es collin\'eaires associ\'ees dans les processus inclusifs.
A l'ordre dominant en log\ $1/x$ et log\ $Q^2$, nous \'ecrivons un syst\`eme
unifi\'e d'\'equations satisfaisant cette contrainte. Cela conduit \`a un
comportement des fonctions de structure \`a petit $x$ moins singulier que
celui pr\'edit par BFKL}
\twocolumn[\maketitle]
\fnm{7}{Talk given in the
Proton and Photon structure and diffractive interaction session
at the Workshop on Deep Inelastic scattering and QCD,
Paris, April 1995}
\section{Introduction}
There exist two different tools in perturbative QCD when one wants to get
predictions for quark and gluon structure functions measured in
 Deep-Inelastic Scattering:
The first one, valid when $Q^2$ is large
, is a resummation to all order of the
leading logarithms of $Q^2$\ ($LLQ^2$), namely terms of the type
 $\alpha_s^n ln(Q^2/\Lambda^2_{QCD})^n$,
due to the {\it collinear} \ singularity of the radiative corrections.
It leads to the so called "Altarelli-Parisi" (or DGLAP)
evolution equations \cite{dglap,rev} at leading order.
The second one, valid when the Bjorken variable $x$
is small, is a resummation to all order of the
leading logarithm of $x$\ ($LL1/x$), namely terms like $\alpha_s^n (ln 1/x)^n$,
due to the {\it infra-red} singularities of the radiative soft part. This
tedious
calculation was performed twenty years ago by L.N.Lipatov and collaborators
(BFKL
equation) and predicts a singular behaviour of the proton structure function
at small $x$ \cite{bfkl}. HERA revived this
result, since the data on the
quark structure function inside proton at very small $x$ are in qualitative
agreement with the BFKL prediction \cite{donnee}.

Since the HERA experiment covers a very large range in $x$ and $Q^2$, it
 would be of great interest to have a unified system of these two equations.
In the past, this possibility has already been discussed. It
was first noticed that a system of equation taking into account both $LL1/x$
and $LLQ^2$ terms can be written down, by a precise combination of the
corresponding integral
kernels \cite {glr}. More recently, it has been proved rigorously that
a unified description of the gluon radiation in the whole $x$-range is
possible,
 due to the property of {\it angular ordering} ~\cite{ccfm}.
In this picture, one can show that the dominant contribution due to collinear
singularities (present in any
gluon production amplitude) comes from the regions satisfying the
following kinematical property:

\begin{equation}
Q/x\gg ...\gg \theta _i\gg \theta _{i-1}\gg ...\gg \theta _1
\label{ord}
\end{equation}

where $\theta _i\approx (q_t)_i/x_i$ are the angles of the emitted gluon with
 respect to the direction of the first emitted gluon momentum. The two previous
regimes are recovered in two different limits:
i) at $x$ of the order of unity, corresponding to non strongly ordered $x_i$,
one recovers the $q_t$ ordering and the $LLQ^2$ resummation \cite{rev}.
ii) at small $x$, the gluon momentum fractions $x_i$ are strongly ordered, and
 the relation (\ref{ord}) doesn't imply $q_t$-ordering. Thus collinear
 singularities
 as well as infrared ones can contribute to $LL1/x$ singularities.

The key point is that for structure functions (this is not the case for
non-inclusive quantities) the collinear singularities cancel in such
a way that
one recovers the BFKL evolution ~\cite{ccfm,m}. We propose here
a scheme of such unification for both glue and quarks, which fullfill this
constraint in an effective way~\cite{pw}.

\section{Unified QCD evolution equations}
\subsection{Equations in Mellin space}
We will restrict ourselves, for the sake of simplicity, to the case of a fixed
coupling constant $\bar {\alpha}_S$, as for the original BFKL derivation. We
consider the double inverse Mellin transform of the singlet ($F_s$) and gluon
($F_G$) structure function with respect to $Q^2$ and $x$:

\begin{eqnarray}
\label{doublemellin}
F_{S,G}(x,Q^2) & = & \int \frac{d\gamma }{2i\pi }\ e^{\gamma \ln Q^2/\Lambda
^2} \nonumber \\
 \times &\int&
\frac{dj}{2i\pi }\ e^{(j-1)\ln 1/x}\varphi _{S,G}(j,\gamma ).
\end{eqnarray}

\noindent The DGLAP equations (for fixed $\bar {\alpha }_S$) can
be written in matrix form for $\varphi _S$ and $\varphi _G$ as follows:

\begin{eqnarray}
\label{systdglap}
\left(
\begin{tabular}{l}
$\varphi _S^{}$ \\
$\varphi _G$
\end{tabular}
\right)
&\equiv&
\left(
\begin{tabular}{l}
$\varphi _S^{(0)}$\\
$\varphi _G^{(0)}$
\end{tabular}
\right) \nonumber
\\
&+& \frac{\overline{\alpha}_S}{4\pi \gamma}%
\left(
\begin{tabular}{ll}
$\nu _F$
&
$2n_F\phi _G^F$
\\
$\phi _F^G$
 &
\ \ \ $\nu _G$
\end{tabular}
\right)
\left(
\begin{tabular}{l}
$\varphi _S^{}$ \\
$\varphi _G$
\end{tabular}
\right),\ \ \
\end{eqnarray}


\noindent where $\left\{ \nu _G,\nu _F,\phi _G^F,\phi _F^G\right\} $ are the
usual $(j$-dependent) DGLAP weights~\cite{rev}, and $\varphi
_{S,G}^{(0)}$ are the initial conditions.

The equation (\ref{systdglap}) must be modified so as to take into account the
BFKL contribution in the gluon sector, due to soft singularities.
This dominant contribution can be expressed as a singularity in the $j-$plane
at the value
\begin{equation}
j_L=1+\frac{\bar {\alpha }_SN_C}\pi \chi \left( \gamma \right)  \label{singul}
\end{equation}

where

\begin{equation}
\chi (\gamma )\equiv 2\psi (1)-\psi (\gamma )-\psi (1-\gamma )\ ; \
\psi (\gamma )\equiv \frac{d\ln \Gamma (\gamma )}{d\gamma }
\label{chi}
\end{equation}

\noindent is the eigenvalue-function of the BFKL kernel. Closing the path of
integration
in $j$ around the rightest pole given
by (\ref{singul}), one gets:

\begin{eqnarray}
\label{domin}
F_{S,G}(x,q^2) & = & \int \frac{d\gamma }{2i\pi }\ e^{\gamma \ln Q^2/\Lambda
^2} \
e^{\bar {\alpha }^S\frac{N_c}\pi \chi (\gamma )\ln 1/x} \nonumber \\
&\simeq & \left( _{%
\overline{\Lambda }^2}^{Q^2}\right) ^{1/2}x^{-\overline{\alpha }_S\frac{N_c}%
\pi4 \ln 2},
\end{eqnarray}

where a saddle point method is used and gives the dominant contribution
at $\gamma _c=1/2$, corresponding to $\chi (\gamma _c)=4\ln 2$.

In order to implement this singular behaviour (\ref{domin}) in
(\ref{systdglap}), we replace the gluonic contribution to the anomalous
dimension by the following \cite{pw}:

\begin{equation}
\nu _{G(j)}\longrightarrow \nu _{G(j)}^{*}=\gamma\ \chi (\gamma )\left\{ \nu
_G+\Psi \right\} -\Psi ,  \label{vg}
\end{equation}

\noindent where $\Psi $ is an arbitrary function holomorphic in the $j$%
-plane near $j=1$ and below.
 Inserted in equation (\ref{systdglap}), this modification
provides a system of unified equations mixing the DGLAP and BFKL
kernels.
Indeed, inverting the relation (3), after the replacement $\nu
_G\rightarrow \nu _G^{*}$, one gets

\begin{eqnarray}
&&\label{unif}
\left(
\begin{tabular}{l}
$\varphi _G^{}$ \\
$\varphi _S$%
\end{tabular}
\right) \equiv \frac 1{D(j,\gamma )} \nonumber \\
&\times& \left(
\begin{tabular}{ll}
\begin{tabular}{l}
$1-\frac{\overline{\alpha }_S}{4\pi \gamma }\ \nu _F^{}$%
\end{tabular}
& $\ \ \ \ \frac{\overline{\alpha }_S}{4\pi \gamma} \ \phi _F^G$ \\
$\frac{\overline{\alpha }_S}{4\pi \gamma }2n_F\ \phi _G^F$ & $1-\frac{%
\overline{\alpha }_S^{}}{4\pi \gamma }\ \nu _G^{*}$%
\end{tabular}
\right) \left(
\begin{tabular}{l}
$\varphi _G^{(0)}$ \\
$\varphi _S^{(0)}$%
\end{tabular}
\right),\ \ \ \ \
\end{eqnarray}

\noindent with
\begin{eqnarray}
\label{denom}
D(j,\gamma )=1 &-& \frac{\overline{\alpha }}{4\pi \gamma }\left(
\nu ^{*}_G+\ \nu _F\right) \nonumber \\
&+& \left( \frac{\overline{\alpha }}{4\pi \gamma }%
\right) ^2\left( \nu _F\nu _G^{*}-2n_F\phi _G^F\phi _F^G\right) .
\end{eqnarray}

Before studying this denominator (\ref{denom}), let us note that in principle
$\phi _F^G$ also gets a contribution from infrared singularities, such
that it should be changed into
$\gamma\ \chi (\gamma )\left\{ \nu
_G+\Psi_1 \right\} -\Psi_1$ where $\Psi_1$ has the same properties as
$\Psi$. This will be discused elsewhere~\cite{pwap}.

The zeroes of $D(j,\gamma)$ depend on the region in the complex $j-$plane
involved in the inverse Mellin transform (\ref{doublemellin}), and thus on
the region in $x$ one is looking at:

i) for $x$ of the order of unity, $\overline{\alpha }_S\ln 1/x\ll 1,$ the
modification (\ref{vg}) has no effect, since the zeroes of $D(j,\gamma )$ are
obtained for small values of $\gamma $ (of order $\overline{\alpha }_S).$ In
that limit, one gets from the definition of $\chi (\gamma ): $

\begin{equation}
\chi (\gamma )\approx 1/\gamma +{\cal O}(\gamma ^2);\ \nu _G^{*}\approx \nu
_G+{\cal O}(\overline{\alpha }^3),  \label{chivgappr}
\end{equation}

\noindent and one recovers the ordinary DGLAP equations~\cite{dglap} (at fixed
$\overline{\alpha }_S).$

\noindent ii) When $\overline{\alpha }_S\ln 1/x={\cal O}(1),$ the singular
structure of the BFKL kernel drives the relevant domain of
the integration over $\gamma $ in (\ref{doublemellin}) near the ''critical''
value $\gamma_c=1/2.$
\noindent One recovers the singular behaviour compatible
with the BFKL calculations. Taking the appropriate limit $j\rightarrow 1,$ $%
\overline{\alpha }_S/(j-1)={\cal O}(1):$

\begin{equation}
\label{bfkllim}
D(j,\gamma )\ \propto \ 1-\frac{\overline{\alpha }N_C \chi (\gamma )}{4\pi
(j-1)}\stackunder{j\rightarrow 1}{\approx}1-\frac{\overline{\alpha }}%
\pi N_C\frac{4\ln 2}{j-1}
\end{equation}

\subsection{Constraint on the $\psi$ function}

The $\psi(j)$ function we have introduced seems to be arbitrary when we try
to unify DGLAP and BFKL equations. In fact, because of the precise
cancellation of collinear singularities that we emphasized in the
introduction, the
collinear singularities arising from quark-loop contribution
should cancel at small-x. Indeed, angular ordering
for radiated quarks (\ref{ord}) can in principle modify the $LL1/x$
singularity. Despite the fact that this cancellation is well established
for gluons, this is not proved when ''finite parts'' are
included~\cite{marfinie}.

We thus impose this cancellation around $j=j_l$, namely, considering
$D(j,\gamma)$ at first order in $\overline{\alpha}_S$ (a
complete discussion will be presented elsewhere \cite{pwap}):
\begin{eqnarray}
\Psi (j_L) &\approx &\nu _F(j_L)  \label{cancel} \\
\nu _G(j_L)+\nu _F(j_L) &=&\left[ \frac{\overline{\alpha }N_C \log 2}\pi
\right] ^{-1}.  \nonumber
\end{eqnarray}

An equivalent constraint is to impose that the conformal properties of the
BFKL kernel~\cite{li} would be preserved, namely that the critical conformal
weight should stay at $\gamma = 1/2$~\cite{pw}. This is a statement at
$LL1/x$ level.

The system of equations (\ref{cancel}) leads to a decrease of the
effective dominant singularity with respect to the BFKL one
\cite{pw}. This is in agreement with HERA data \cite{donnee}.

Note that the energy-momentum conservation can be implemented in the system
of equations (\ref{unif}) in a consistant way~\cite{pw}. It imposes that
$\psi(2) = \nu_G(2) = 2$. This constraint is not related to the
previous one at $j_l$ and is not responsible in our model for the
decrease of the $j-$plane singularity (in contrary to the discussion of
Ref. \cite{ekl}).

\section{Conclusion}
 The unification of the evolution equations for structure functions is
possible, combining the leading-logarithmic contributions in both $x$ and
$Q^2$ variables at fixed $\overline{\alpha}_S$. The constraint due to the
mixing of $LL1/x$ and $LLQ^2$ leads to a shift down of the $LL1/x$
prediction.

It would be interesting to know the phenomenological consequences of our
system of unified equations.
Work is in progress in order to take also into account the running of
$\alpha_S$ and the next-leading order terms.
\begin{center}
{\large\bf Aknowledgements}
\end{center}
All these results come from a fruitful collaboration with R. Peschanski.
It is a pleasure to thank L.N.Lipatov for discussion during the Cambridge
meeting
about conformal
properties of unified equations.

\end{document}